\documentclass[12pt,aps,prd,showpacs,amsmath,amssymb]{revtex4}
\usepackage{graphicx}
\usepackage{epstopdf}
\usepackage{multirow}
\usepackage{subfigure}
\usepackage{appendix}
\input epsf
\begin{document}
\title{The magnetic moment of $P_{c}(4312)$ as a $\bar{D}\Sigma_{c}$ molecular state}
\author{Yong-Jiang Xu$^{1}$\footnote{xuyongjiang13@nudt.edu.cn}, Yong-Lu Liu$^1$, and Ming-Qiu Huang$^{1,2}$\footnote{corresponding author: mqhuang@nudt.edu.cn}}
\affiliation{$^1$Department of Physics, College of Liberal Arts and Sciences, National University of Defense Technology , Changsha, 410073, Hunan, China}
\affiliation{$^2$Synergetic Innovation Center for Quantum Effects and Applications, Hunan Normal University, Changsha,  410081, Hunan, China}
\date{}
\begin{abstract}
In this paper, we tentatively assign the $P_{c}(4312)$ to be a $\bar{D}\Sigma_{c}$ molecular state with quantum number $J^{P}=\frac{1}{2}^{-}$, and calculate its magnetic moment using the QCD sum rule method in external weak electromagnetic field. Starting with the two-point correlation function in external electromagnetic field and expanding it in power of the electromagnetic interaction Hamiltonian, we extract the magnetic moment from the linear response to the external electromagnetic field. The numerical value of the magnetic moment of $P_{c}(4312)$ is $\mu_{P_{c}}=1.75^{+0.15}_{-0.11}$.
\end{abstract}
\pacs{11.25.Hf,~ 11.55.Hx,~ 13.40.Gp.} \maketitle

\section{Introduction}\label{sec1}

In Ref.\cite{lhcb1}, LHCb collaboration reported the discoveries of two pentaquark states $P_{c}(4380)$ and $P_{c}(4450)$ in the $J/\psi p$ invariant mass spectrum of the process $\Lambda_{b}\rightarrow J/\psi pK$. In 2019, they confirmed the $P_{c}(4450)$ state consisting of two narrow overlapping peaks $P_{c}(4440)$ and $P_{c}(4457)$, and observed a new narrow pentaquark state $P_{c}(4312)$ \cite{lhcb2}. Following these experimental discoveries, there have been many theoretical studies concerning these pentaquark states through various models/methods, such as the meson-baryon molecular scenario \cite{molecular1,molecular2,molecular3,molecular4,molecular5,molecular6,molecular7,molecular8,molecular9,molecular10,
molecular11,molecular12,molecular13,molecular14,molecular15,molecular16,molecular17,molecular18,molecular19,molecular20,
molecular21,molecular22,molecular23,molecular24,molecular25,molecular26,molecular27}, the compact five quark states \cite{pentaquark1,pentaquark2,pentaquark3,pentaquark4,pentaquark5,pentaquark6,pentaquark7,pentaquark8,pentaquark9,
pentaquark10,pentaquark11,pentaquark12,pentaquark13,pentaquark14,pentaquark15,pentaquark16,pentaquark17}, kinematical triangle singularity \cite{triangle} and so on.

In Ref.\cite{molecular27}, we assumed the $P_{c}(4312)$ as a $\bar{D}\Sigma_{c}$ molecular state with quantum number $\frac{1}{2}^{-}$, and studied the decay of $P_{c}(4312)$ to $J/\psi p$ and to $\eta_{c}p$ with the QCD sum rule method. The QCD sum rule method \cite{SVZ} is a nonperturbative analytic formalism firmly entrenched in QCD with minimal modeling and has been successfully applied in almost every aspect of strong interaction physics. In Ref.\cite{Balitsky,Ioffe1,Ioffe2}, the QCD sum rule method was extended to calculate the magnetic moments of the nucleon and hyperon in the external field method. In this method, a static electromagnetic field is introduced which couples to the quarks and polarizes the QCD vacuum, and the magnetic moments of hadrons can be extracted from the linear response to this field. Later, a more systematic studies was made for the magnetic moments of the octet baryons \cite{octet1,octet2,octet3,octet4}, the decuplet baryons \cite{decuplet1,decuplet2,decuplet3,decuplet4} and the $\rho$ meson \cite{rho}. In Refs.\cite{wangzhigang} and \cite{xuyongjiang}, the authors calculated the magnetic moment of $Z_{c}(3900)$ as an axialvector tetraquark state and an axialvector molecular state, respectively.

In the present work, we extend this method to the investigation of the magnetic moment of the $P_{c}(4312)$ state viewed as a $\bar{D}\Sigma_{c}$ molecular state with quantum number $J^{P}=\frac{1}{2}^{-}$. Electromagnetic multipole moments are the major and meaningful parameters of hadrons. Analysis of the electromagnetic multipole moments of the exotic states can help us get valuable knowledge about the electromagnetic properties of these states, the charge distributions inside them, their charge radius and geometric shapes and finally their internal substructures.

The rest of the paper is organized as follows. In Sec. \ref{sec2}, the sum rule for the magnetic moment of the $P_{c}(4312)$ state is given. Sec. \ref{sec3} is devoted to the numerical analysis and a short summary is given in section \ref{sec4}. In Appendix \ref{appendix2}, the spectral densities are shown.

\section{The derivation of the sum rules}\label{sec2}

The starting point of our calculation is the time-ordered correlation function in the QCD vacuum in the presence of a constant background electromagnetic field $F_{\mu\nu}$,
\begin{equation}\label{2-point correlator}
\Pi(p)=i\int dx^{4}e^{ipx}\langle0\mid\textsl{T}[J^{P_{c}}(x)\bar{J}^{P_{c}}(0)]\mid0\rangle_{F}
=\Pi^{(0)}(p)+\Pi^{(1)}_{\mu\nu}(p)F^{\mu\nu}+\cdots,
\end{equation}
where
\begin{equation}\label{Pc interpotating current}
J^{P_{c}}(x)=[\bar{c}(x)i\gamma_{5}d(x)][\epsilon^{abc}(u^{t}_{a}(x)C\gamma_{\mu}u_{b}(x))\gamma^{\mu}\gamma_{5}c_{c}(x)],
\end{equation}
is the interpolating current of $P_{c}(4312)$ considered as a $\bar{D}\Sigma_{c}$  molecular state with $J^{P}=\frac{1}{2}^{-}$
with $t$ denoting the matrix transposition on the Dirac spinor indices, $C$ meaning charge conjugation operator, and $a, b, c$ being color indices. In the present work, we shall consider the linear response term, $\Pi^{(1)}_{\mu\nu}(p)F^{\mu\nu}$, from which the magnetic moment will be extracted.

The external electromagnetic field can interact directly with the quarks inside the hadron and also polarize the QCD vacuum. As a consequence, the vacuum condensates involved in the operator product expansion of the correlation function in the external electromagnetic field $F_{\mu\nu}$ are,
\begin{itemize}
  \item{dimension-2 operator},
  \begin{equation}
  F_{\mu\nu},
  \end{equation}
  \item{dimension-3 operator},
  \begin{equation}
  \langle0|\bar{q}\sigma_{\mu\nu}q|0\rangle_{F},
  \end{equation}
  \item{dimension-5 operators},
  \begin{equation}
  \langle0|\bar{q}q|0\rangle F_{\mu\nu}, \langle0|\bar{q}g_{s}G_{\mu\nu}q|0\rangle_{F}, \epsilon_{\mu\nu\alpha\beta}\langle0|\bar{q}g_{s}G^{\alpha\beta}q|0\rangle_{F},
  \end{equation}
  \item{dimension-6 operators},
  \begin{equation}
  \langle0|\bar{q}q|0\rangle\langle0|\bar{q}\sigma_{\mu\nu}q|0\rangle_{F}, \langle0|g^{2}_{s}GG|0\rangle F_{\mu\nu},\cdots,
  \end{equation}
  \item{dimension-7 operators},
  \begin{equation}
  \langle0|g^{2}_{s}GG|0\rangle\langle0|\bar{q}\sigma_{\mu\nu}q|0\rangle_{F}, \langle0| g_{s}\bar{q}\sigma\cdot Gq|0\rangle F_{\mu\nu}, \cdots,
  \end{equation}
  \item{dimension-8 operators},
  \begin{eqnarray}
  &&\langle0|\bar{q}q|0\rangle^{2}F_{\mu\nu}, \langle0| g_{s}\bar{q}\sigma\cdot Gq|0\rangle\langle0|\bar{q}\sigma_{\mu\nu}q|0\rangle_{F}, \langle0|\bar{q}q|0\rangle\langle0|\bar{q}g_{s}G_{\mu\nu}q|0\rangle_{F},\nonumber\\ &&\epsilon_{\mu\nu\alpha\beta}\langle0|\bar{q}q|0\rangle\langle0|\bar{q}g_{s}G^{\alpha\beta}q|0\rangle_{F},\cdots,
  \end{eqnarray}
\end{itemize}
and so on. The new vacuum condensates induced by the external electromagnetic field $F_{\mu\nu}$ can be described by introducing new parameters, $\chi$, $\kappa$ and $\xi$, called vacuum susceptibilities as follows,
\begin{eqnarray}
&&\langle0|\bar{q}\sigma_{\mu\nu}q|0\rangle_{F}=ee_{q}\chi\langle0|\bar{q}q|0\rangle F_{\mu\nu},\nonumber\\
&&\langle0|\bar{q}g_{s}G_{\mu\nu}q|0\rangle_{F}=ee_{q}\kappa\langle0|\bar{q}q|0\rangle F_{\mu\nu},\nonumber\\
&&\epsilon_{\mu\nu\alpha\beta}\langle0|\bar{q}g_{s}G^{\alpha\beta}q|0\rangle_{F}=iee_{q}\xi\langle0|\bar{q}q|0\rangle F_{\mu\nu}.
\end{eqnarray}

In order to express the two-point correlation function (\ref{2-point correlator}) physically, we expand it in powers of the electromagnetic interaction Hamiltonian $H_{int}=-ie\int d^{4}yj^{em}_{\alpha}(y)A^{\alpha}(y)$,
\begin{eqnarray}\label{expansion}
\Pi(p)=&&i\int dx^{4}e^{ipx}\langle0\mid\textsl{T}[J^{P_{c}}(x)\bar{J}^{P_{c}}(0)]\mid0\rangle\nonumber\\&&+i\int dx^{4}e^{ipx}\langle0\mid\textsl{T}\{J^{P_{c}}(x)[-ie\int d^{4}yj^{em}_{\mu}(y)A^{\mu}(y)]\bar{J}^{P_{c}}(0)\}\mid0\rangle+\cdots,
\end{eqnarray}
where $j^{em}_{\mu}(y)$ is the electromagnetic current and $A^{\mu}(y)$ is the electromagnetic four-vector.

Inserting two complete sets of physical intermediate states with the same quantum numbers as the current operator $J^{P_{c}}(x)$ into the second term of (\ref{expansion}), we have
\begin{eqnarray}\label{complete}
\Pi(p)=&&e\int d^{4}xd^{4}y\frac{d^{4}k}{(2\pi)^{4}}\frac{d^{4}k^{\prime}}{(2\pi)^{4}}e^{ipx}A^{\mu}(y)\sum_{P_{c},P^{\prime}_{c}}\sum_{s,s^{\prime}}
\frac{-i}{k^{2}-m^{2}_{P_{c}}}\frac{-i}{k^{\prime2}-m^{2}_{P^{\prime}_{c}}}\nonumber\\
&&\langle0|J^{P_{c}}(x)|P^{\prime}_{c}(k^{\prime},s^{\prime})\rangle\langle P^{\prime}_{c}(k^{\prime},s^{\prime})|j^{em}_{\mu}(y)|P_{c}(k,s)\rangle\langle P_{c}(k,s)|\bar{J}^{P_{c}}(0)|0\rangle,
\end{eqnarray}
where the sum $\sum_{P_{c},P^{\prime}_{c}}$ is over all possible intermediate states including the ground state $P_{c}(4312)$ we are interested in, higher resonances and continuum. One can translate the coordinates of the operators in (\ref{complete}) to the origin, carry out the integrals over $x$ and $k^{\prime}$ and then finds that
\begin{eqnarray}
\Pi(p)=&&-e\int d^{4}y\frac{d^{4}k}{(2\pi)^4}e^{i(p-k)y}A^{\mu}(y)\sum_{P_{c},P^{\prime}_{c}}\sum_{s,s^{\prime}}
\frac{1}{(p^{2}-m^{2}_{P^{\prime}_{c}})(k^{2}-m^{2}_{P_{c}})}\nonumber\\
&&\langle0|J^{P_{c}}(0)|P^{\prime}_{c}(p,s^{\prime})\rangle\langle P^{\prime}_{c}(p,s^{\prime})|j^{em}_{\mu}(0)|P_{c}(k,s)\rangle\langle P_{c}(k,s)|\bar{J}^{P_{c}}(0)|0\rangle.
\end{eqnarray}
The sum $\sum_{P_{c},P^{\prime}_{c}}$ can be divided into three parts, the ground-ground term, the ground-excited (continuum) term and the excited (continuum)-excited (continuum) term.

After standard manipulation, the ground-ground term can be written as
\begin{eqnarray}\label{hadronic side}
\Pi^{(1)}_{\mu\nu}(p)F^{\mu\nu}=&&-\frac{\lambda^{2}_{P_{c}}}{4(p^{2}-m^{2}_{P_{c}})^{2}}[2m_{P_{c}}\mu_{P_{c}}\sigma^{\mu\nu}
+\frac{\mu_{P_{c}}-1}{m_{P_{c}}}(p^{2}-m^{2}_{P_{c}})\sigma^{\mu\nu}+\mu_{P_{c}}(\not\!{p}\sigma^{\mu\nu}+
\sigma^{\mu\nu}\not\!{p})\nonumber\\&&+2i\frac{\mu_{P_{c}}-1}{m_{P_{c}}}(p^{\mu}\gamma^{\nu}-p^{\nu}\gamma^{\mu})\not\!{p}]F_{\mu\nu},
\end{eqnarray}
where we make use of the following formulas,
\begin{equation}
\langle0\mid J^{P_{c}}(0)\mid P_{c}(p,s)\rangle=\lambda_{P_{c}} u(p,s),
\end{equation}
and
\begin{equation}
\langle P^{\prime}_{c}(k^{\prime},s^{\prime})|j^{em}_{\mu}(0)|P_{c}(k,s)\rangle=\bar{u}(k^{\prime},s^{\prime})[F_{1}(Q^{2}) \gamma_{\mu}+F_{2}(Q^{2})i\sigma_{\mu\nu}\frac{q^{\nu}}{2m_{P_{c}}}]u(k,s),
\end{equation}
with $q=k^{\prime}-k$ and $Q^{2}=-q^{2}$. $\lambda_{P_{c}}$ and $u(k,s)$ are the pole residue and Dirac spinor of the $P_{c}(4312)$ state, respectively. The Lorentz invariant form factors $F_{1}(Q^{2})$ and $F_{2}(Q^{2})$ are related to the charge and magnetic form factors by
\begin{eqnarray}
&&G_{C}(Q^{2})=F_{1}(Q^{2})-\frac{Q^{2}}{4m^{2}_{P_{c}}}F_{2}(Q^{2}),\nonumber\\
&&G_{M}(Q^{2})=F_{1}(Q^{2})+F_{2}(Q^{2}).
\end{eqnarray}
The magnetic moment $\mu_{P_{c}}$ is given by $G_{M}(0)$.

Now, we consider the ground-excited (continuum) and the excited (continuum)-excited (continuum) parts. For each Lorentz structure in (\ref{hadronic side}), the hadronic representation of the the second term in (\ref{expansion}) is
\begin{equation}
\frac{A_{P_{c}P_{c}}}{(p^{2}-m^{2}_{P_{c}})^{2}}+\sum_{P^{*}_{c}}\frac{A_{P_{c}P^{*}_{c}}}
{(p^{2}-m^{2}_{P_{c}})(p^{2}-m^{2}_{P^{*}_{c}})}+\sum_{P^{*}_{c}}\frac{A_{P^{*}_{c}P^{*}_{c}}}
{(p^{2}-m^{2}_{P^{*}_{c}})^{2}},
\end{equation}
where $A_{P_{c}P_{c}}$, $A_{P_{c}P^{*}_{c}}$ and $A_{P^{*}_{c}P^{*}_{c}}$ are constants, and the symbol $\sum_{P^{*}_{c}}$ means the sum over the excited states and the integral over continuum. The first term in the above equation is the ground state pole which contains the desired magnetic moment $\mu_{P_{c}}$. The second term represents the transition between the ground sate and the excited states (continuum) induced by the external electromagnetic field. The last term is the contributions from pure excited states (continuum). Making Borel transform, one has
\begin{equation}
\frac{A_{P_{c}P_{c}}}{M^{2}_{B}}e^{-m^{2}_{P_{c}}/M^{2}_{B}}+e^{-m^{2}_{P_{c}}/M^{2}_{B}}[\sum_{P^{*}_{c}}\frac{A_{P_{c}P^{*}_{c}}}
{m^{2}_{P^{*}_{c}}-m^{2}_{P_{c}}}(1-e^{-(m^{2}_{P^{*}_{c}}-m^{2}_{P_{c}})/M^{2}_{B}})]
+\sum_{P^{*}_{c}}\frac{A_{P^{*}_{c}P^{*}_{c}}}{M^{2}_{B}}e^{-m^{2}_{P^{*}_{c}}/M^{2}_{B}},
\end{equation}
where $M^{2}_{B}$ is the Borel parameter. It is obvious that the transition between the ground sate and the excited states (continuum) gives a contribution which is not suppressed exponentially relative to the ground state. We can approximate the quantity in the square brackets by a constant. The third term is suppressed exponentially relative to the ground state and can be modeled in the usual way by introducing the continuum model and threshold parameter.

On the other hand, $\Pi(p)$ can be calculated theoretically via OPE method at the quark-gluon level. To this end, one can substitute the interpolating current $J^{P_{c}}(x)$ (\ref{Pc interpotating current}) into the correlation function (\ref{2-point correlator}), contract the relevant quark fields by Wick's theorem and find
\begin{eqnarray}
\Pi^{OPE}(p)=&&-2i\epsilon_{abc}\epsilon_{a^{\prime}b^{\prime}c^{\prime}}\int d^{4}x e^{ipx}\{\gamma^{\mu}\gamma_{5}S^{(c)}_{cc^{\prime}}(x)\gamma^{\nu}\gamma_{5}\nonumber\\&&Tr[(i\gamma_{5})S^{(d)}_{dd^{\prime}}(x)(i\gamma_{5})S^{(c)}_{d^{\prime}d}(-x)]
Tr[\gamma_{\mu}S^{(u)}_{bb^{\prime}}(x)\gamma_{\nu}CS^{(u)T}_{aa^{\prime}}(x)C]\}_{F},
\end{eqnarray}
where $S^{(c)}(x)$ and $S^{(q)}(x), q=u, d$ are the full charm- and up (down)-quark propagators, whose expressions are given in Appendix \ref{appendix1}. Through dispersion relation, $\Pi^{OPE}(p)$ can be written as
\begin{eqnarray}\label{QCD side}
\Pi^{OPE}(p)=&&\sigma^{\mu\nu}F_{\mu\nu}\int^{\infty}_{4m^{2}_{c}}ds\frac{\rho_{1}(s)}{s-p^2}+ (\not\!{p}\sigma^{\mu\nu}+\sigma^{\mu\nu}\not\!{p})F_{\mu\nu}\int^{\infty}_{4m^{2}_{c}}ds\frac{\rho_{2}(s)}{s-p^2}
\nonumber\\&&+i(p^{\mu}\gamma^{\nu}-p^{\nu}\gamma^{\mu})\not\!{p}F_{\mu\nu}\int^{\infty}_{4m^{2}_{c}}ds\frac{\rho_{3}(s)}{s-p^2}
+\cdots,
\end{eqnarray}
where $\rho_{i}(s)=\frac{1}{\pi}\mbox{Im}\Pi^{OPE}_{i}(s), i=1,2,3$ are the spectral densities. We will choose the Lorentz structure $i(p^{\mu}\gamma^{\nu}-p^{\nu}\gamma^{\mu})\not\!{p}F_{\mu\nu}$ to obtain our sum rule for the magnetic moment $\mu_{P_{c}}$ because of its better convergence. The spectral density $\rho_{3}(s)$ is given in Appendix \ref{appendix2}.

Finally, with the help of the quark-hadron duality and the above discussion, we match the phenomenological side (\ref{hadronic side}) and the QCD representation (\ref{QCD side}) for the Lorentz structure $i(p^{\mu}\gamma^{\nu}-p^{\nu}\gamma^{\mu})\not\!{p}F_{\mu\nu}$
\begin{equation}
-\frac{\lambda^{2}_{P_{c}}}{2(p^{2}-m^{2}_{P_{c}})^{2}}\frac{\mu_{P_{c}}-1}{m_{P_{c}}}+\frac{a}{m^{2}_{P_{c}}-p^{2}}
+\int^{\infty}_{s^{P_{c}}_{0}}ds\frac{\rho_{3}(s)}{s-p^2}+\mbox{subtractions}=\int^{\infty}_{4m^{2}_{c}}ds\frac{\rho_{3}(s)}{s-p^2},
\end{equation}
where the constant $a$ is introduced to parameterize the contributions of the ground-excited states (continuum) transition and $s^{P_{c}}_{0}$ is the threshold parameter.
Subtracting the contributions of pure excited states (continuum), one gets
\begin{equation}
-\frac{\lambda^{2}_{P_{c}}}{2(p^{2}-m^{2}_{P_{c}})^{2}}\frac{\mu_{P_{c}}-1}{m_{P_{c}}}+\frac{a}{m^{2}_{P_{c}}-p^{2}}
+\mbox{subtractions}=\int^{s^{P_{c}}_{0}}_{4m^{2}_{c}}ds\frac{\rho_{3}(s)}{s-p^2}.
\end{equation}
In order to eliminate the subtractions, it is necessary to make a Borel transform which can also improve the convergence of the OPE series and suppress the contributions from the excited and continuum states. As a result, we have
\begin{equation}\label{magnetic moment sum rule1}
(-\frac{\mu_{P_{c}}-1}{2m_{P_{c}}M^{2}_{B}}+A)\lambda^{2}_{P_{c}}e^{-m^{2}_{P_{c}}/M^{2}_{B}}=
\int^{s^{P_{c}}_{0}}_{4m^{2}_{c}}dse^{-s/M^{2}_{B}}\rho_{3}(s),
\end{equation}
where $A=\frac{a}{\lambda^{2}_{P_{c}}}$ and $M^{2}_{B}$ is the Borel parameter.

\section{Numerical analysis}\label{sec3}

The input parameters needed in numerical analysis are presented in Table \ref{input parameters}. For the vacuum susceptibilities $\chi$, $\kappa$ and $\xi$, we take the values $\chi=-(3.15\pm0.30)\mbox{GeV}^{-2}$, $\kappa=-0.2$ and $\xi=0.4$ determined in the detailed QCD sum rules analysis of the photon light-cone distribution amplitudes \cite{P.Ball}. Besides these parameters, we should determine the working intervals of the threshold parameter $s^{0}_{P_{c}}$ and the Borel mass $M^{2}_{B}$ in which the magnetic moment is stable. The continuum threshold is related to the square of the mass of the first exited states having the same quantum number as the interpolating field and we use the value determined in Ref.\cite{molecular27}, while the Borel parameter is determined by demanding that both the contributions of the higher states and continuum are sufficiently suppressed and the contributions coming from higher dimensional operators are small.
\begin{table}[htb]
\caption{Some input parameters needed in the calculations.}\label{input parameters}
\begin{tabular}{|c|c|}
  \hline
  Parameter      &   Value    \\
  \hline
  {$\langle\bar{q}q\rangle$}  &      $-(0.24\pm0.01)^{3}\mbox{GeV}^{3}$                     \\
  {$\langle g_{s}\bar{q}\sigma Gq\rangle$} & $(0.8\pm0.1)\langle\bar{q}q\rangle \mbox{GeV}^{2}$ \\
  {$\langle g^{2}_{s}GG\rangle$}    &     $0.88\pm0.25\mbox{GeV}^{4}$                \\
  {$m_{c}$}  &    $1.275^{+0.025}_{-0.035}\mbox{GeV}$\cite{M.Tanabashi}                   \\
  {$m_{P_{c}}$}  &   $4311.9\pm0.7^{+6.8}_{-0.6}\mbox{MeV}$\cite{lhcb2}     \\
  $\lambda_{P_{c}}$  &   $1.91^{+0.12}_{-0.13}\times10^{-3}\mbox{GeV}^{6}$\cite{molecular27}    \\
  \hline
\end{tabular}
\end{table}

We define two quantities, the ratio of the pole contribution to the total contribution (RP) and the ratio of the highest dimensional term in the OPE series to the total OPE series (RH), as followings,
\begin{eqnarray}
&&RP\equiv\frac{\int^{s^{P_{c}}_{0}}_{4m^{2}_{c}}ds\rho_{3}(s)e^{-\frac{s}{M^{2}_{B}}}}{\int^{\infty}_{4m^{2}_{c}}ds\rho_{3}(s)e^{-\frac{s}{M^{2}_{B}}}},
\nonumber\\&&RH\equiv\frac{\int^{s^{P_{c}}_{0}}_{4m^{2}_{c}}ds\rho^{(d=9)}_{3}(s)e^{-\frac{s}{M^{2}_{B}}}}{\int^{s^{P_{c}}_{0}}_{4m^{2}_{c}}ds\rho_{3}(s)e^{-\frac{s}{M^{2}_{B}}}}.
\end{eqnarray}

\begin{figure}[htb]
\subfigure[]{
\includegraphics[width=7cm]{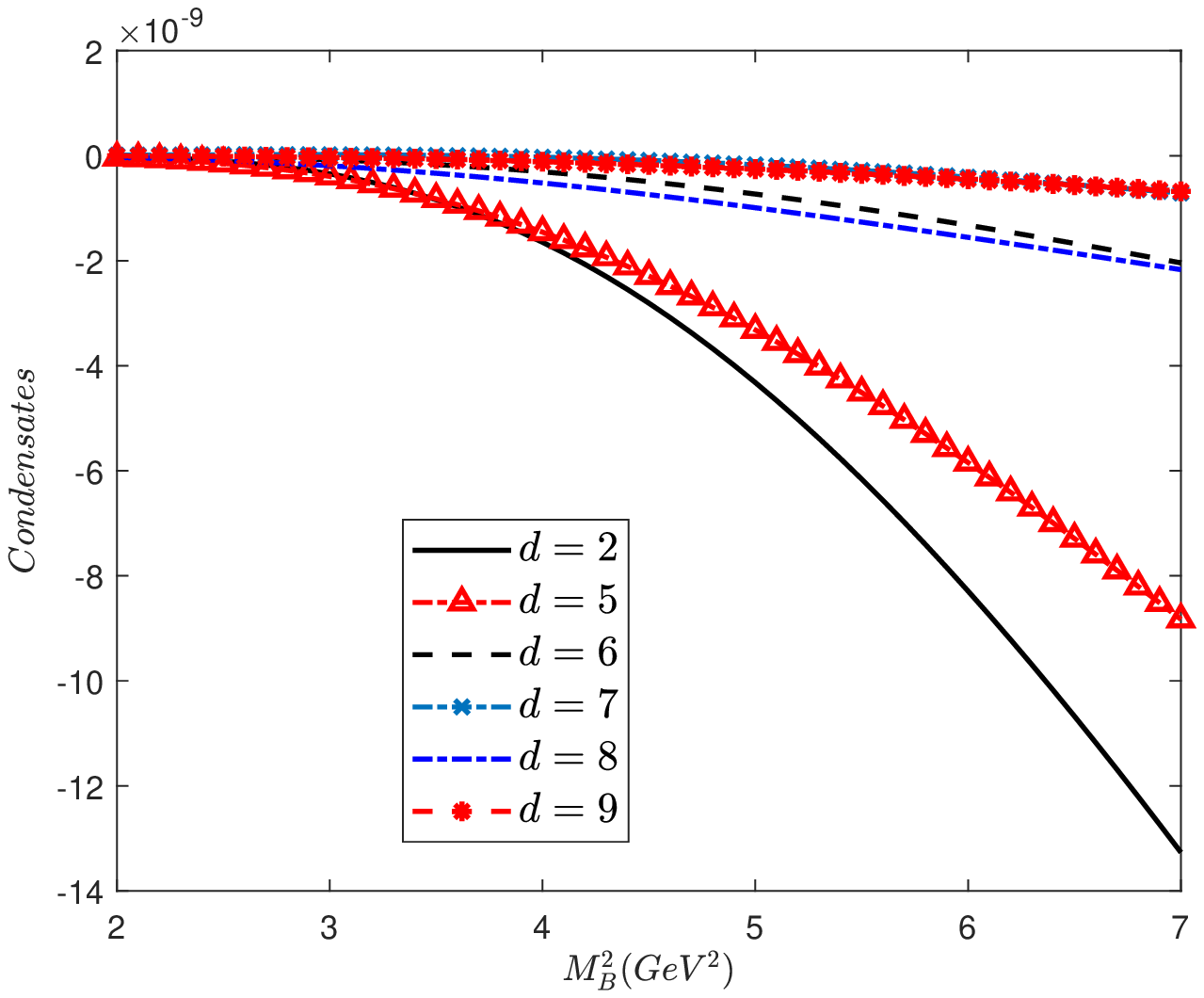}}
\subfigure[]{
\includegraphics[width=7cm]{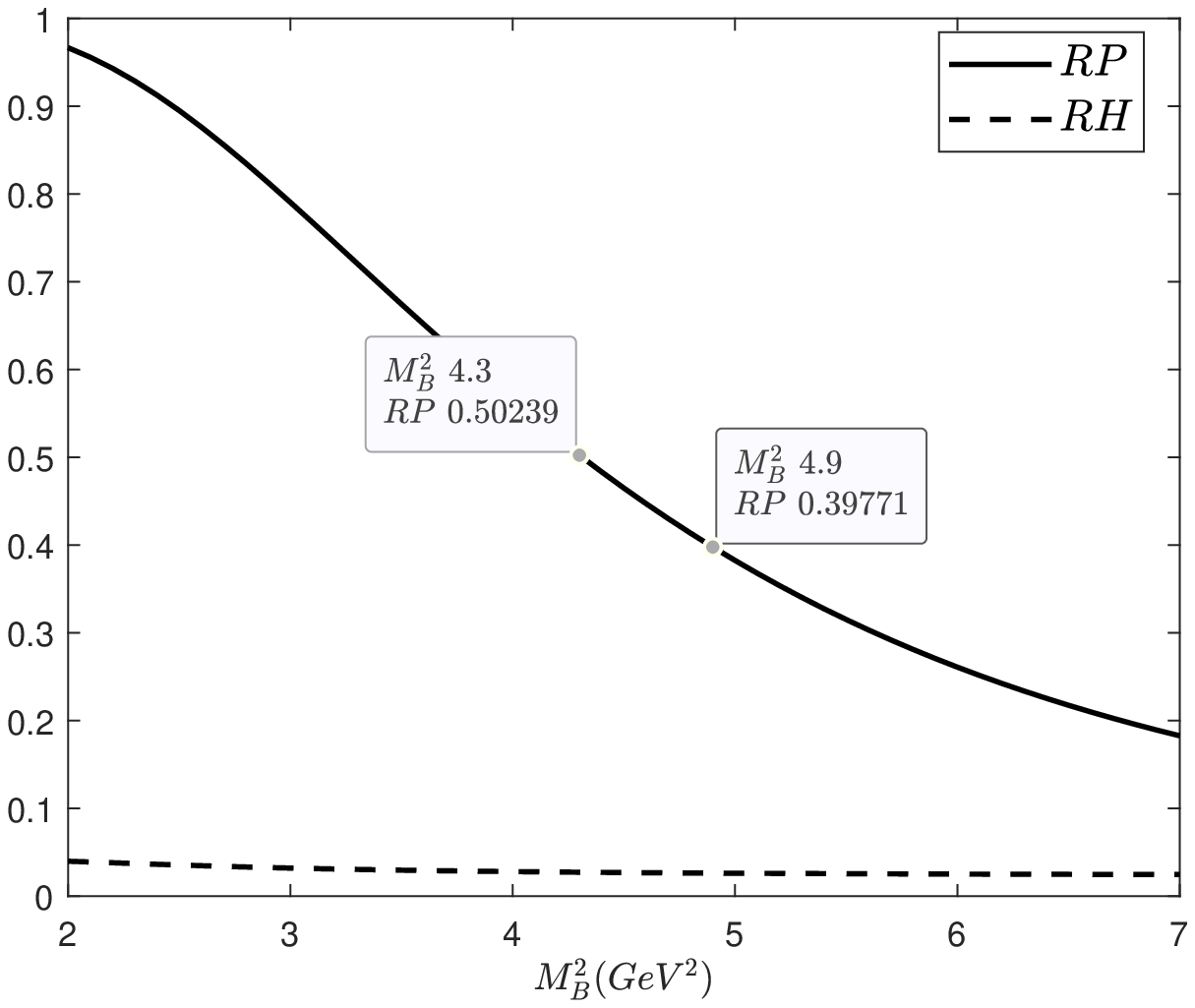}}
\caption{(a) denotes the various condensates as functions of $M^{2}_{B}$ with $\sqrt{s^{P_{c}}_{0}}=4.8\mbox{GeV}$; (b) represents RP and RH varying with $M^{2}_{B}$ at $\sqrt{s^{P_{c}}_{0}}=4.8\mbox{GeV}$.}\label{MB_range}
\end{figure}

Firstly, we determine the working region of the $M^{2}_{B}$. In Fig.\ref{MB_range}(a), we compare the various OPE contributions as functions of $M^{2}_{B}$ with $\sqrt{s^{P_{c}}_{0}}=4.8\mbox{GeV}$. From it one can see that the OPE has good convergence. Fig.\ref{MB_range}(b) shows RP and RH varying with $M^{2}_{B}$ at $\sqrt{s^{P_{c}}_{0}}=4.8\mbox{GeV}$. The figure shows that the requirement $RP\geq50\%$ ($RP\geq40\%$) gives $M^{2}_{B}\leq4.3\mbox{GeV}^{2}$ ($M^{2}_{B}\leq4.9\mbox{GeV}^{2}$).

\begin{figure}[htb]
\subfigure[]{
\includegraphics[width=7cm]{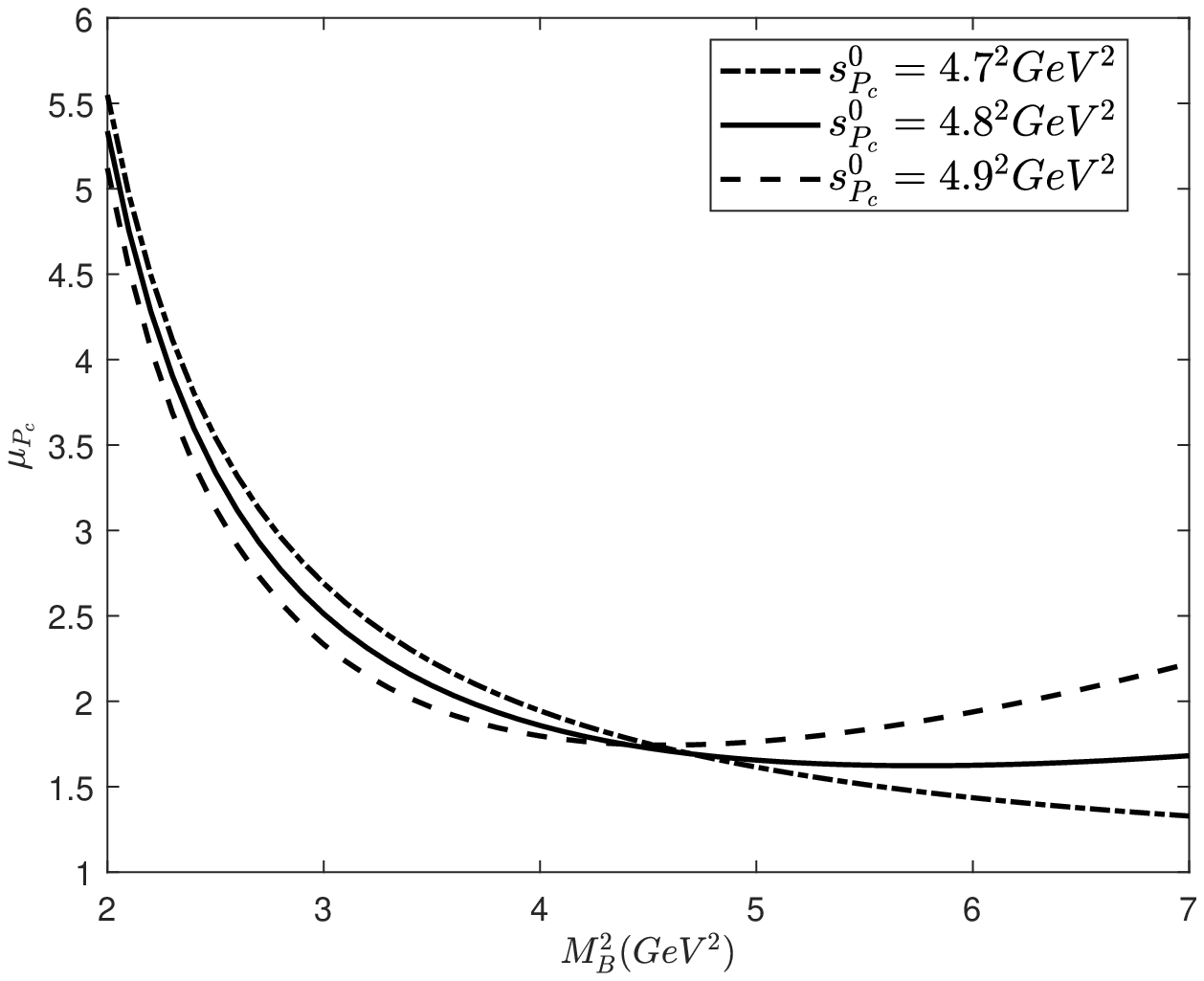}}
\subfigure[]{
\includegraphics[width=7cm]{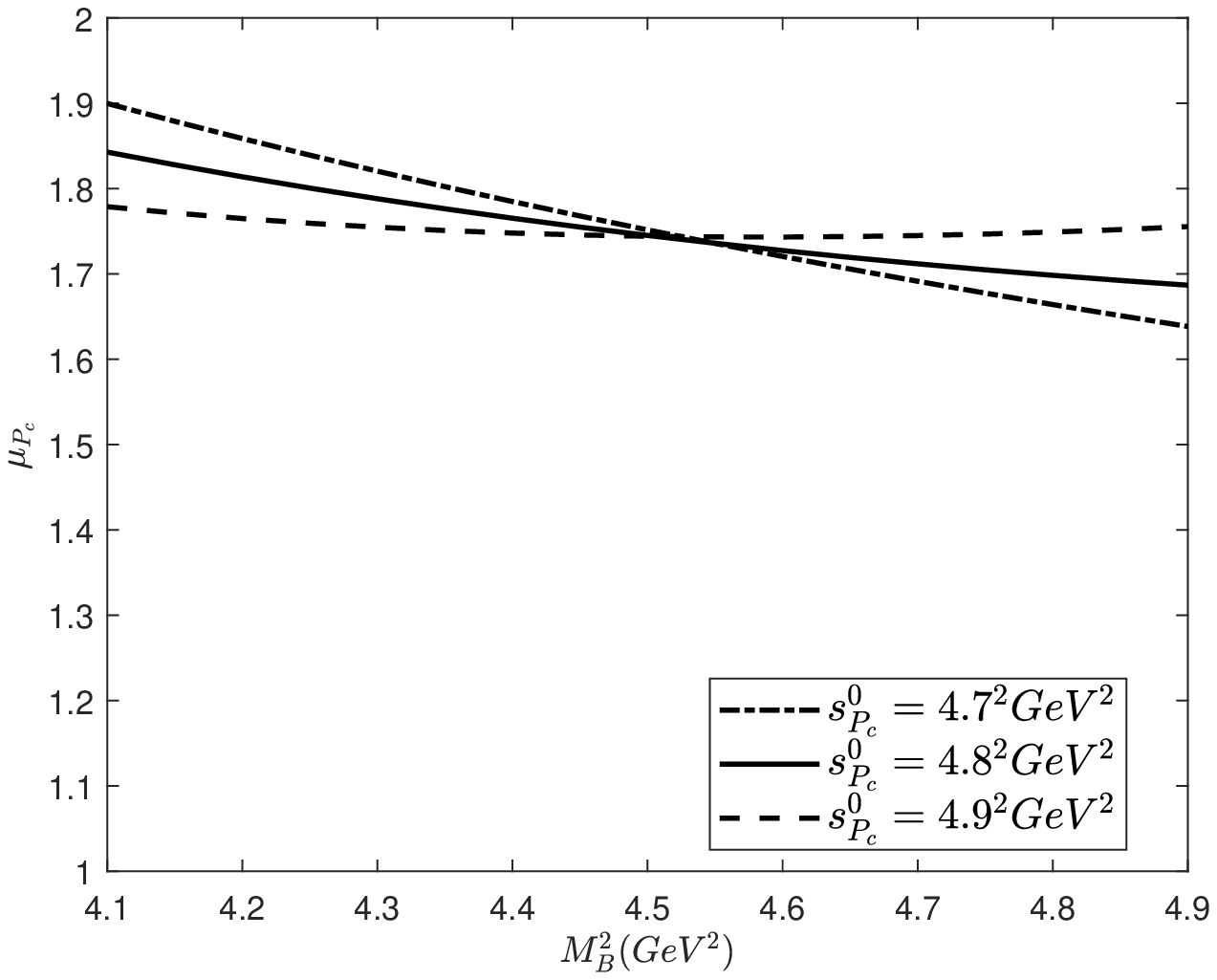}}
\caption{The dependence of the magnetic moment $\mu_{P_{c}}$ on $M^{2}_{B}$ at three different values of $s^{P_{c}}_{0}$.}\label{magneticmoment}
\end{figure}

Fig.\ref{magneticmoment}(a) shows the dependence of the magnetic moment $\mu_{P_{c}}$ on the Borel mass $M^{2}_{B}$ in the interval of $2\mbox{GeV}^{2}\leq M^{2}_{B} \leq 7\mbox{GeV}^{2}$. From the figure we can see that $\mu_{P_{c}}$ depends strongly on $M^{2}_{B}$ and $s^{P_{c}}_{0}$ as $4\mbox{GeV}^{2}\leq M^{2}_{B}$ or $M^{2}_{B}\geq5\mbox{GeV}^{2}$. In order to have a larger working interval of the Borel mass $M^{2}_{B}$, we require $RP\geq40\%$. As a result, we limit $M^{2}_{B}$ from $4.1\mbox{GeV}^{2}$ to $4.9\mbox{GeV}^{2}$. The result is shown in Fig.\ref{magneticmoment}(b), from which we can read reliably the value of the magnetic moment, $\mu_{P_{c}}=1.75^{+0.15}_{-0.11}$. Because the hadron's magnetic moments contain important information on the charge distributions inside them, their geometric shapes and finally their quark configuration. This value can be confronted to the experimental data in the future and tell us whether it is reasonable to view the $P_{c}(4312)$ as a $\bar{D}\Sigma_{c}$ molecular state with quantum number $J^{P}=\frac{1}{2}^{-}$.

\section{Conclusion}\label{sec4}

In this paper, we tentatively assign the $P_{c}(4312)$ to be a $\bar{D}\Sigma_{c}$ molecular state with quantum number $J^{P}=\frac{1}{2}^{-}$, calculate its magnetic moment using the QCD sum rule method in the external weak electromagnetic field. Starting with the two-point correlation function in the external electromagnetic field and expanding it in power of the electromagnetic interaction Hamiltonian, we extract the magnetic moment from the linear response to the external electromagnetic field. The numerical value of the magnetic moment of the $P_{c}(4312)$ state is $\mu_{P_{c}}=1.75^{+0.15}_{-0.11}$. The prediction can be confronted to the experimental data in the future and give important information about the inner structure of the $P_{c}(4312)$ state.

\acknowledgments  This work was supported by the National
Natural Science Foundation of China under Contract No.11675263.

\begin{appendix}
\section{The quark propagators}\label{appendix1}
The full quark propagators are given as
\begin{eqnarray}
 S^{q}_{ij}(x)=&&\frac{i \not\!{x}}{2\pi^{2}x^4}\delta_{ij}-\frac{m_{q}}{4\pi^2x^2}\delta_{ij}-\frac{\langle\bar{q}q\rangle}{12}\delta_{ij}
 +i\frac{\langle\bar{q}q\rangle}{48}m_{q}\not\!{x}\delta_{ij}-\frac{x^2}{192}\langle g_{s}\bar{q}\sigma Gq\rangle \delta_{ij}\nonumber\\
 &&+i\frac{x^2\not\!{x}}{1152}m_{q}\langle g_{s}\bar{q}\sigma Gq\rangle \delta_{ij}-i\frac{g_{s}t^{a}_{ij}G^{a}_{\mu\nu}}{32\pi^2x^2}(\not\!{x}\sigma^{\mu\nu}+\sigma^{\mu\nu}\not\!{x})\nonumber\\
 &&+i\frac{\delta_{ij}e_{q}F_{\mu\nu}}{32\pi^2x^2}(\not\!{x}\sigma^{\mu\nu}+\sigma^{\mu\nu}\not\!{x})
 -\frac{\delta_{ij}e_{q}\chi\langle\bar{q}q\rangle\sigma^{\mu\nu}F_{\mu\nu}}{24}\nonumber\\
 &&+\frac{\delta_{ij}e_{q}\langle\bar{q}q\rangle F_{\mu\nu}}{288}(\sigma^{\mu\nu}-2\sigma^{\alpha\mu}x_{\alpha}x^{\nu})\nonumber\\
 &&+\frac{\delta_{ij}e_{q}\langle\bar{q}q\rangle F_{\mu\nu}}{576}[(\kappa+\xi)\sigma^{\mu\nu}x^{2}-(2\kappa-\xi)\sigma^{\alpha\mu}x_{\alpha}x^{\nu}]+\cdots
 \end{eqnarray}
 for light quarks, and
 \begin{eqnarray}
 S^{Q}_{ij}(x)=i\int\frac{d^{4}k}{(2\pi)^4}e^{-ikx}&&[\frac{\not\!{k}+m_{Q}}{k^2-m^{2}_{Q}}\delta_{ij}
 -\frac{g_{s}t^{a}_{ij}G^{a}_{\mu\nu}}{4}\frac{\sigma^{\mu\nu}(\not\!{k}+m_{Q})+(\not\!{k}+m_{Q})\sigma^{\mu\nu}}
 {(k^2-m^{2}_{Q})^{2}}\nonumber\\
 &&+\frac{\langle g^{2}_{s}GG\rangle}{12}\delta_{ij}m_{Q}\frac{k^2+m_{Q}\not\!{k}}{(k^2-m^{2}_{Q})^{4}}\nonumber\\
 &&+\frac{\delta_{ij}e_{Q}F_{\mu\nu}}{4}\frac{\sigma^{\mu\nu}(\not\!{k}+m_{Q})+(\not\!{k}+m_{Q})\sigma^{\mu\nu}}
 {(k^2-m^{2}_{Q})^{2}}+\cdots]
 \end{eqnarray}
 for heavy quarks. In these expressions $t^{a}=\frac{\lambda^{a}}{2}$ and $\lambda^{a}$ are the Gell-Mann matrix, $g_{s}$ is the strong interaction coupling constant, and $i, j$ are color indices, $e_{Q(q)}$ is the charge of the heavy (light) quark and $F_{\mu\nu}$ is the external electromagnetic field.

\section{The spectral densities}\label{appendix2}
In this appendix, we will give the explicit expression of the spectral density $\rho_{3}(s)$.
\begin{equation}
\rho_{3}(s)=\rho^{(d=2)}_{3}+\rho^{(d=3)}_{3}(s)+\rho^{(d=5)}_{3}(s)+\rho^{(d=6)}_{3}(s)
+\rho^{(d=7)}_{3}(s)+\rho^{(d=8)}_{3}(s)+\rho^{(d=9)}_{3}(s),
\end{equation}
with
\begin{eqnarray}
\rho^{(d=2)}_{3}(s)=&&-\frac{m_{c}}{12288\pi^{8}}\int^{a_{max}}_{a_{min}}da\int^{1-a}_{b_{min}}db\frac{1}{a^{2}b^{4}}(1-a-b)^{4}(m^{2}_{c}(a+b)-abs)^{3}\nonumber\\
&&+\frac{m_{c}}{3072\pi^{8}}\int^{a_{max}}_{a_{min}}da\int^{1-a}_{b_{min}}db\frac{1}{a^{2}b^{3}}(1-a-b)^{3}(m^{2}_{c}(a+b)-abs)^{3},
\end{eqnarray}
\begin{equation}
\rho^{(d=3)}_{3}(s)=0,
\end{equation}
\begin{eqnarray}
\rho^{(d=5)}_{3}(s)=&&\frac{m^{2}_{c}\langle0|\bar{q}q0|\rangle}{384\pi^{6}}\int^{a_{max}}_{a_{min}}da\int^{1-a}_{b_{min}}db\frac{1}{ab^{2}}(1-a-b)^{3}(m^{2}_{c}(a+b)-abs)\nonumber\\
&&-\frac{m^{2}_{c}\langle0|\bar{q}q|0\rangle}{128\pi^{6}}\int^{a_{max}}_{a_{min}}da\int^{1-a}_{b_{min}}db\frac{1}{ab}(1-a-b)^{2}(m^{2}_{c}(a+b)-abs),
\end{eqnarray}
\begin{eqnarray}
\rho^{(d=6)}_{3}(s)=&&-\frac{m^{3}_{c}\langle0| g^{2}_{s}GG|0\rangle}{147456\pi^{8}}\int^{a_{max}}_{a_{min}}da\int^{1-a}_{b_{min}}db\frac{1}{ab^{2}}(1-a-b)^{4}\nonumber\\
&&-\frac{m_{c}\langle0| g^{2}_{s}GG|0\rangle}{12288\pi^{8}}\int^{a_{max}}_{a_{min}}da\int^{1-a}_{b_{min}}db\frac{1}{ab^{2}}(1-a-b)^{3}(m^{2}_{c}(a+b)-abs)\nonumber\\
&&-\frac{m_{c}\langle0| g^{2}_{s}GG|0\rangle}{24576\pi^{8}}\int^{a_{max}}_{a_{min}}da\int^{1-a}_{b_{min}}db\frac{1}{b^{2}}(1-a-b)^{2}(m^{2}_{c}(a+b)-abs)\nonumber\\
&&+\frac{m_{c}\langle0| g^{2}_{s}GG|0\rangle}{12288\pi^{8}}\int^{a_{max}}_{a_{min}}da\int^{1-a}_{b_{min}}db\frac{1}{b^{3}}(1-a-b)^{3}(m^{2}_{c}(a+b)-abs)\nonumber\\
&&+\frac{m_{c}\langle0| g^{2}_{s}GG|0\rangle}{8192\pi^{8}}\int^{a_{max}}_{a_{min}}da\int^{1-a}_{b_{min}}db\frac{1}{ab^{2}}(1-a-b)^{2}(a+2b)(m^{2}_{c}(a+b)-abs)\nonumber\\
&&+\frac{m^{3}_{c}\langle0| g^{2}_{s}GG|0\rangle}{36864\pi^{8}}\int^{a_{max}}_{a_{min}}da\int^{1-a}_{b_{min}}db\frac{1}{b^{3}}(1-a-b)^{3}(a+b),
\end{eqnarray}
\begin{eqnarray}
\rho^{(d=7)}_{3}(s)=&&\frac{m^{2}_{c}\langle0|g_{s}\bar{q}\sigma\cdot Gq0|\rangle}{512\pi^{6}}\int^{a_{max}}_{a_{min}}da\int^{1-a}_{b_{min}}db\frac{1}{b}(1-a-b)^{2}\nonumber\\
&&-\frac{m^{2}_{c}\langle0|g_{s}\bar{q}\sigma\cdot Gq0|\rangle}{768\pi^{6}}\int^{a_{max}}_{a_{min}}da\int^{1-a}_{b_{min}}db\frac{1}{ab}(1-a-b)^{3}\nonumber\\
&&-\frac{m^{2}_{c}\langle0|g_{s}\bar{q}\sigma\cdot Gq0|\rangle}{256\pi^{6}}\int^{a_{max}}_{a_{min}}da\int^{1-a}_{b_{min}}db(1-a-b)\nonumber\\
&&+\frac{m^{2}_{c}\langle0|g_{s}\bar{q}\sigma\cdot Gq0|\rangle}{256\pi^{6}}\int^{a_{max}}_{a_{min}}da\int^{1-a}_{b_{min}}db\frac{1}{b}(1-a-b)^{2},
\end{eqnarray}
\begin{eqnarray}
\rho^{(d=8)}_{3}(s)=&&-\frac{m_{c}(2\kappa-\xi)\langle0|\bar{q}q|0\rangle^{2}}{576\pi^{4}}\int^{a_{max}}_{a_{min}}da\int^{1-a}_{b_{min}}dba\nonumber\\
&&-\frac{m_{c}\langle0|\bar{q}q|0\rangle^{2}}{144\pi^{4}}\int^{a_{max}}_{a_{min}}da\int^{1-a}_{b_{min}}dba,
\end{eqnarray}
\begin{eqnarray}
\rho^{(d=9)}_{3}(s)=&&\frac{m^{4}_{c}\langle0|\bar{q}q|0\rangle\langle0| g^{2}_{s}GG|0\rangle}{27648\pi^{6}M^{2}_{B}}\int^{a_{max}}_{a_{min}}da\frac{(1-a-b_{min})^{3}}{(as-m^{2}_{c})b^{2}_{min}}\nonumber\\
&&-\frac{m^{2}_{c}\langle0|\bar{q}q|0\rangle\langle0| g^{2}_{s}GG|0\rangle}{9216\pi^{6}}\int^{a_{max}}_{a_{min}}da\frac{(1-a-b_{min})^{3}}{(as-m^{2}_{c})b_{min}}\nonumber\\
&&-\frac{m^{2}_{c}\langle0|\bar{q}q|0\rangle\langle0| g^{2}_{s}GG|0\rangle}{9216\pi^{6}}\int^{a_{max}}_{a_{min}}da\frac{a(1-a-b_{min})}{(as-m^{2}_{c})}\nonumber\\
&&-\frac{m^{4}_{c}\langle0|\bar{q}q|0\rangle\langle0| g^{2}_{s}GG|0\rangle}{4608\pi^{6}M^{2}_{B}}\int^{a_{max}}_{a_{min}}da\frac{a(1-a-b_{min})^{2}}{(as-m^{2}_{c})b^{2}_{min}}\nonumber\\
&&+\frac{m^{2}_{c}\langle0|\bar{q}q|0\rangle\langle0| g^{2}_{s}GG|0\rangle}{1536\pi^{6}}\int^{a_{max}}_{a_{min}}da\frac{a(1-a-b_{min})^{2}}{(as-m^{2}_{c})b_{min}}\nonumber\\
&&+\frac{m^{2}_{c}\langle0|\bar{q}q|0\rangle\langle0| g^{2}_{s}GG|0\rangle}{3072\pi^{6}}\int^{a_{max}}_{a_{min}}da\frac{a(1-a-b_{min})}{(as-m^{2}_{c})}.
\end{eqnarray}
In the above equations, $a_{max}=\frac{1+\sqrt{1-\frac{4m^{2}_{c}}{s}}}{2}$, $a_{min}=\frac{1-\sqrt{1-\frac{4m^{2}_{c}}{s}}}{2}$ and $b_{min}=\frac{am^{2}_{c}}{as-m^{2}_{c}}$.

\end{appendix}


\end{document}